\documentstyle[preprint,eqsecnum,aps,epsf]{revtex}
\tightenlines
\begin{document}
\title
{Semiclassical density of degeneracies in quantum regular systems}
\author{A. J. Fendrik and M. J. S\'anchez}
\maketitle
\noindent
\begin{center}
\center{\it Departamento de F\'{\i}sica  J. J. Giambiagi,\\
 Facultad de Ciencias Exactas  y Naturales, \\
Universidad de Buenos Aires.
Ciudad Universitaria, 1428 Buenos Aires, Argentina.}\\ 
\end{center}
\begin{abstract}
The spectrum of eigenenergies of a quantum integrable system whose hamiltonian
depends on a single parameter shows degeneracies (crossings) when the parameter
varies. We derive a semiclassical expression for the density of crossings
in the plane energy-parameter, that is the number of crossings per unit of energy and unit of parameter, in terms of classical periodic orbits. We compare the results of the semiclassical formula with exact quantum calculations for two specific quantum integrable billiards.  
\end{abstract}

\section*{Introduction}
\label{int}
The analysis of energy spectra of different physical bounded systems has always 
been an interesting subject in quantum mechanics. The energy level spacings
as well as the existence of degeneracies (crossings) in such systems have been
widely studied during recent years. The number of degrees of freedom, the 
separability of the problem and the number of free parameters involved, among 
others, are very important features which have to be taken into account when
a given quantum spectrum is analyzed.

Starting from Percival's ideas \cite{parc} for systems with more than one degree
of freedom, two kinds of spectra have been distinguished. The regular spectrum,
whose level spacing is characterized by a Poisson distribution \cite{betab}, is
associated with integrable problems (the harmonic oscillator is an exception).
The other one, the irregular spectrum closely characterized by 
spectral statistic of the Gaussian ensembles, corresponds to non-integrable 
systems \cite{pechu,bohi,yuka}.
This statistical behavior can be related to crossings and repulsions between levels with the same symmetry when the spectra are analyzed as a function of
the parameters of the problem. In fact, integrable systems depending on one parameter exhibit many crossings while non-integrable systems show repulsions and double-hyperbola curves (avoided crossings), rather than degeneracies. 
So, for one parameter dependent systems, the presence of crossings or avoided 
crossings are quantum fingerprints of the properties of the classical dynamics.
When the system depends on two parameters, the surfaces of energy (that define 
in such cases the eigenenergies) will cross in continuum curves leading to 
diedric intersections when the system remains integrable or they will intersect 
at isolated points (``diabolical points'') if the systems is not integrable 
(assuming that it has time reversal symmetry)\cite{berry2}.

Distributions of avoided crossings (for one parameter dependent systems) 
according to properties such that the closest approach or the mean and
difference between the slopes of the involved levels were already established for a generic quantum system employing parametric random-matrix 
models \cite{wil,wil1}.  The same models, extended to two and more parameters, were used to study distributions of diabolical points \cite{wil3} and other
relevant distributions of singular points in the spectra \cite{walk} . 

On the other hand, it is well known that classical periodic orbits are essential elements to develop semiclassical quantization methods. Not only for integrable 
systems (through the Berry-Tabor formula \cite{betab2} ) but also for the non-
integrable ones (using the Gutzwiller methods \cite{gutzwi}) the spectral density can be formally described in terms of closed orbits of the classical system. Therefore, it would not be surprising that other quantum densities such as densities of degeneracies or densities of avoided crossings can be related to the periodic orbits. 

In the present paper, we find the density of degeneracies as an expansion in
terms of the periodic orbits for classical integrable systems depending on one parameter.     

In section \ref{densi} we introduce the density of degeneracies. The
semiclassical version can be written as a sum of a smooth part and 
oscillating contributions depending on the periodic orbits of the classical
system. Section \ref{suave} is devoted to compute the smooth part of
the density of degeneracies and related distributions for
two specific integrable systems whose hamiltonians depend on a single parameter
in a different functional form.
We study the rectangular billiard of sides $a$ and $b$ where the parameter is the ratio $\mu=b/a$ (shape parameter) and the Aharonov-Bohm cylindrical billiard where the parameter is the magnetic flux, that is $\mu=\phi$. 
The oscillating contributions are computed in Section \ref{oscpart}.
Finally Section \ref{cr} is devoted to concluding remarks.
We have included an Appendix that contains the appropriate derivations of the
semiclassical density of degeneracies when the quantum system has not Kramer's degeneracies (like the Aharonov-Bohm cylindrical billiard).

\section{The density of degeneracies}
\label{densi}

We consider an integrable system whose hamiltonian depends on a single parameter
$\mu$.
That is:
\begin{equation}
H=H(\vec{I},\mu)
\end{equation}
where $\vec{I} \equiv (I_{1},..,I_{n})$ and $I_{i}=\frac{1}{2\pi} \oint p_{i} dq_{i}$
are the action variables. To obtain the eigenvalues, we can 
employ the E.B.K. semiclassical quantization rule 
\begin{equation}
I_{i}=\hbar (n_{i}+\alpha_{i}/4) \; ,
\end{equation}
where $n_{i}= 0,1,2,...$ and $\alpha_{i}$ are the Maslov index \cite{gutzwi}.
So, we establish the quantum eigenenergies $E$ by
\begin{equation}
E(\vec{n},\mu)=H(\vec{n}+\vec{\alpha}/4,\mu) \; .
\end{equation}
When we consider the eigenenergies as a function of the
parameter $\mu$, the spectrum shows degeneracies (crossings).
They occur whenever
\begin{equation}
H(\vec{n}+\vec{\alpha}/4,\mu)-H(\vec{n}^{\prime}+\vec{\alpha}/4,\mu)=0 \;. 
\end{equation}
Given $\vec{n}$ and $\vec{n}^{\prime}$, this equation determines
the values of the parameter $\mu$ for which the eigenenergies labeled by
$\vec{n}$ and $\vec{n}^{\prime}$ are degenerated, 
\begin{equation}
\label{para}
\mu=L(\vec{n}+\vec{\alpha}/4,\vec{n}^{\prime}+\vec{\alpha}/4) \; .
\end{equation} 
We define the density of degeneracies  $\rho_{c}(E,\mu)$ as the number of 
crossings that occurs in the energy interval $[E,E+dE]$ and in the parameter
interval $[\mu,\mu+d\mu]$.

Therefore, using the E.B.K. rule, we can write $\rho_{c}(E,\mu)$ as follows:
\begin{equation} 
\label{rho}
\rho_{c}(E,\mu)=\frac{1}{2} \sum_{\vec{n}} \sum_{\vec{n}^{\prime}} 
\delta(E-H(\vec{n}+\vec{\alpha}/4,\mu)) \delta(\mu-L(\vec{n}+
\vec{\alpha}/4,\vec{n}^{\prime}+\vec{\alpha}/4)) \; ,
\end{equation}
where we disregard as in Ref.\cite{betab} possible degeneracy factors. 
Employing the Poisson sumation formula in 
Eq. (\ref{rho})
we write, 
\begin{eqnarray}
\label{rhosum}
\rho_{c}(E,\mu) & = &\frac{1}{2\hbar^{2n}}\sum_{\vec{m}}\sum_{\vec{m}^{\prime}}
\exp{\left[-i \frac{\pi}{2}(\vec{\alpha} \cdot \vec{m} + \vec{\alpha} \cdot 
\vec{m}^{\prime})\right]}  \nonumber  \\ 
 &  & \times \int_{\vec{I} \geq 0} \int_{\vec{I}^{\prime} \geq 0} 
d^{n}I \; d^{n}I^{\prime}
\delta\left( E-H(\vec{I},\mu) \right) \delta\left(\mu-L(\vec{I},\vec{I}^{\prime})\right)
\exp{\left[ i \frac{2 \pi}{\hbar} (\vec{m} \cdot \vec{I} + \vec{m}^{\prime} \cdot
\vec{I}^{\prime} )\right]} \; . 
\end{eqnarray}

In the following, to make the expressions more handled, we assume two degrees
of freedom. 

To eliminate the $\delta$-functions, we change the integration variables
as follows,
\begin{equation}
(I_{1},I_{1}^{\prime},I_{2},I_{2}^{\prime}) \rightarrow
(I_{1},I_{1}^{\prime},\xi_{1},\xi_{2}) \; ,
\end{equation}
where we have defined
\begin{eqnarray}
\xi_{1} &\equiv& E-H(I_{1},I_{2},\mu) \nonumber \; , \\
\xi_{2} &\equiv& \mu-L(I_{1},I_{2},I_{1}^{\prime},I_{2}^{\prime}) \; .
\end{eqnarray}
Therefore
\begin{equation}
dI_{1}dI_{2}dI_{1}^{\prime}dI_{2}^{\prime}=
\frac{1}{\left| \frac{\partial{H}}{\partial{I_{2}}}
 \frac{\partial{L}}{\partial{I_{2}^{\prime}}} 
\right |} dI_{1}dI_{2} d\xi_{1}d\xi_{2} \; .
\end{equation}
The partial derivative $\partial{L}/\partial{I_{2}^{\prime}}$
is defined by the implicit equation
\begin{equation}
H(I_{1},I_{2},L)-H(I_{1}^{\prime},I_{2}^{\prime},L)=0
\end{equation}
and we obtain 
\begin{equation}
\left |\frac{\partial{L}}{\partial{I_{2}^{\prime}}} \right|=
\frac{1}{ \left | \left.\frac{\partial{H}}{\partial{L}} \right)_{I} 
-\left. \frac{\partial{H}}{\partial{L}} \right)_{I^{\prime}} \right| }
\frac{\partial{H}}{\partial{I_{2}^{\prime}}} \; ,
\end{equation}
where $\frac{\partial{H}}{\partial{L}})_{I} \; \left(\frac{\partial{H}}{\partial{L}} )_{I^{\prime}}\right)$ is the partial derivative evaluated in $I \; \left(I^{\prime}\right)$.
After integration over $\xi_{1}$ and $\xi_{2}$ follows 
\begin{eqnarray}
\label{2d}
\rho_{c}(E,\mu)&=&\frac{1}{2 \hbar^{4}} \sum_{\vec{m}} \sum_{\vec{m}^{\prime}}
\exp{\left[-i \frac{\pi}{2} (\vec{\alpha} \cdot \vec{m} + \vec{\alpha} \cdot 
\vec{m}^{\prime})\right]} \nonumber \\
& & \times \int_{I_{1} \geq 0} \int_{I_{1}^{\prime} \geq 0} 
dI_{1} dI_{1}^{\prime}
\frac {\left| \partial_{\mu}{H}-\partial_{\mu}{H^{\prime}} \right|}
{\omega_{2} \omega_{2}^{\prime}} 
\exp{\left[ i \frac{2 \pi}{\hbar} (\vec{m} \cdot \vec{I} + \vec{m}^{\prime} \cdot
\vec{I}^{\prime} )\right]} \; ,
\end{eqnarray}
where 
\begin{eqnarray}
\partial_{\mu}{H} & \equiv &
\frac{\partial{H\left(I_{1},I_{2}(E,I_{1}),\mu\right)}}{\partial{\mu}} \;,\\
\partial_{\mu}{H^{\prime}} & \equiv &
\frac{\partial{H\left(I_{1}^{\prime},I_{2}^{\prime}(E,I_{1}^{\prime}),
\mu\right)}}{\partial{\mu}} \; , \\ 
\omega_{2} & \equiv & \frac{\partial{H(I_{1},I_{2},\mu)}}
{\partial {I_{2}}} \; ,\\
\omega_{2}^{\prime} & \equiv & \frac{\partial{H(I_{1}^{\prime},I_{2}^{\prime},
\mu)}}{ \partial{I_{2}}^{\prime} } \; .
\end{eqnarray}

The first term in the Eq.(\ref{2d}) (with 
$m_{1}=m_{2}=m_{1}^{\prime}=m_{2}^{\prime}=0$) corresponds to replace in
expression Eq.(~\ref{rho})  the
quadruple sum by the quadruple integral. It is a smooth function (i.e. non
oscillating) of $E$ and $\mu$, 
\begin{equation}
\label{smooth}
\left< \rho_{c}(E,\mu) \right>=\frac{1}{2 \hbar^{4}} \int_{I_{1}} \int_{I_{1}^{\prime}}
dI_{1} dI_{1}^{\prime} \frac {\left| \partial_{\mu}{H}-\partial_{\mu}{H^{\prime}} \right|}
{\omega_{2} \; {\omega_{2}}^{\prime}} \; .
\end{equation}
Starting from Eq.(~\ref{smooth}) we can determine other relevant distributions. As an example, we can obtain the smooth part of the distribution of crossings $\left<\rho_{c}(E,\mu,V)\right>$
according to the difference between the slopes of the levels,
\begin{equation}
V \equiv \left| \partial_{\mu}{H}-\partial_{\mu}{H^{\prime}} \right| \; . 
\end{equation}
To obtain such distribution (in the following DCDS), we perform the change of the integration variables in Eq.(~\ref{smooth}) 
\begin{equation}
(I_{1},I_{1}^{\prime}) \rightarrow (I_{1},V) \; .
\end{equation}
Therefore
\begin{equation}
dI_{1}dI_{1}^{\prime}= \frac{dI_{1}dV}
{\left|\partial_{\mu}{\omega_{1}^{\prime}}-(\omega_{1}^{\prime}/
\omega_{2}^{\prime})\partial_{\mu}{\omega_{2}^{\prime}} \right|} \; ,
\end{equation}
where the $\omega$'s must be considered as a function of $I_{1}, V$ and
$\mu$. Leaving out the integration over $V$, follows
\begin{equation}
\label{emuv}
\left<\rho_{c}(E,\mu,V)\right>= V \int_{I_{1}} \frac{dI_{1}}{\omega_{2}
\left|\omega_{2}^{\prime}
\partial_{\mu}{\omega_{1}^{\prime}} - \omega_{1}^{\prime}   \partial_{\mu}{\omega_{2}^{\prime}} \right|} \; .
\end{equation}
We stress that the dependence on $V$ in Eq.(\ref{emuv}) is not only given
by the prefactor $V$ but also by the integrand and the limits of integration in the integral over $I_{1}$. 

The other terms in Eq.(\ref{2d}) contain oscillating
functions and we utilize the stationary phase technique to evaluate them.   
The conditions of stationary phase lead to the periodicity conditions, namely
\begin{eqnarray}
\label{perio}
\frac {\omega_{1}}{\omega_{2}} & = &\frac {m_{1}}{m_{2}} \;, \\
\frac {\omega_{1}^{\prime}}{\omega_{2}^{\prime}} & = & 
\frac {m_{1}^{\prime}}{m_{2}^{\prime}} \; .
\end{eqnarray}

Because $\omega_{k}$'s ($\omega_{k}^{\prime}$'s) are positive, the sums over
$m_{k}$'s ($m_{k}^{\prime}$'s) in Eq.(\ref{2d}) are restricted to the first and the third quadrants. The numbers $m_{1},m_{2}$ define the topology of the periodic
orbits. However, there could be pairs of periodic orbits that 
having the same topology, they are related to each other through time reversal
transformation (non-self retracing orbits).

For the terms $m_{1}\neq 0, m_{2}\neq 0, m_{1}^{\prime} = m_{2}^{\prime}= 0$
($m_{1}=m_{2}=0, m_{1}^{\prime} \neq 0 , m_{2}^{\prime} \neq 0$) we perform
first the integration over the variable $I_{1}^{\prime}\;(I_{1})$ and next we 
evaluate the second integral using the stationary phase approximation.  
Finally, we obtain
\begin{equation}
\label{osmain}
\rho_{c}(E,\mu)_{osc1} = \frac{1}{ \hbar^{7/2}} 
\sum_{m_{1},m_{2} \neq (0,0)} 
\frac{\Delta_{m_{1},m_{2}}} {\left| m_{2} \frac{d^{2}I_{2}}{d I_{1}^{2}}
\right |^{1/2} } \left[  \int_{I_{1}^{\prime} \geq 0} 
P(I_{1}^{o},I_{1}^{\prime})
dI_{1}^{\prime} \right ] \cos{\left(S(m_{1},m_{2})/\hbar +\theta(m_{1},m_{2})\right)}
\end{equation}
where we have defined
\begin{eqnarray}
P(X,Y) & \equiv & 
\left | \frac{\partial_{\mu}{H}(X)-\partial_{\mu}{H^{\prime}}(Y)}
{ \omega_{2}(X) \omega_{2}(Y)} \right| \; , \\
\label{action} S(m_{1},m_{2}) & \equiv & 2 \pi (m_{1} I_{1}^{o} + m_{2} I_{2}^{o}) \; ,\\ 
\label{theta} \theta(m_{1},m_{2})
& \equiv & \frac{\pi}{4} \mbox{sig} \left(m_{2} \frac{d^{2}I_{2}}{dI_{1}^{2}} 
\right)-\frac{\pi}{2}(\alpha_{1} m_{1} + \alpha_{2} m_{2}) \; ,\\
\Delta_{m_{1},m_{2}} & \equiv & \left\{ \begin{array}{ll}
2    & \mbox{if there are two orbits of topology $m_{1},m_{2}$} \\
1      & \mbox{otherwise} \; .
\end{array}
\right.
\end{eqnarray}
$I_{1}^{o}=I_{1}^{o}(m_{1},m_{2}),I_{2}^{o}=I_{2}^{o}(m_{1},m_{2}) $ 
are the values of the actions given by Eq. (\ref{perio}). 
On the other hand, for $m_{1} \neq 0$ or $m_{2} \neq 0$   
and $m_{1}^{\prime} \neq 0$ or $m_{2}^{\prime} \neq 0$ 
we obtain
\begin{eqnarray}
\label{gral}
\rho_{c}(E,\mu)_{osc2} & = & \frac{2}{\hbar^{3}}
\sum_{m_{1},m_{2}\neq (0,0)} \; \sum_{m_{1}^{\prime},m_{2}^{\prime}\neq(0,0)} 
\frac{\Delta_{m_{1},m_{2}} \Delta_{m_{1}^{\prime},m_{2}^{\prime}} 
P\left(I_{1}^{o}(m_{1},m_{2}),I_{1}^{\prime o}(m_{1}^{\prime},m_{2}^{\prime})\right)}
{\left |  m_{2} m_{2}^{\prime} \frac{d^{2}I_{2}}{dI_{1}^{2}} 
\frac{d^{2}I_{2}^{\prime}} {dI_{1}^{\prime 2} } \right | ^{1/2} } \nonumber \\
&   & \times \cos{\left(S(m_{1},m_{2})/\hbar + \theta(m_{1},m_{2})\right)} 
\cos{\left(S(m_{1}^{\prime},m_{2}^{\prime})/\hbar + 
\theta^{\prime}(m_{1}^{\prime},m_{2}^{\prime})\right)} \; .
\end{eqnarray}

\section{The smooth part of the density of crossings}
\label{suave}
In the present section we compute the smooth part of the density of
crossing starting from expression Eq.(\ref{smooth}) for two specific systems.
We will consider two kinds of billiards whose hamiltonian depend on the
parameter $\mu$ in a different way. 

\subsection{Rectangular billiards}

At first we study the well known rectangular billiard. That is a spinless
particle in a two dimensional rectangular box of sides $a$ and $b$.
The hamiltonian in terms of the action variables is
\begin{equation}
\label{hamilr}
H(I_{1},I_{2})= \frac{\pi^{2}}{2m} \left ( \frac{I_{1}^{2}}{a^{2}} +
\frac{I_{2}^{2}}{b^{2}} \right ) \; .
\end{equation}
In this case we will consider the crossings as a
function of the shape parameter $\mu=b/a$. We fix the area of the box
$A=ab$ as a constant to conserve invariant the smooth part of the density of states.
So the hamiltonian can be written as
\begin{equation}
\label{renor}
h = \mu I_{1}^{2}+ \frac{I_{2}^{2}}{\mu} \; ,
\end{equation}
where $h \equiv \frac{2mAH}{\pi^{2}}$. In the following we will use the hamiltonian given by Eq.(\ref{renor}) and we call the corresponding
energy as $\epsilon \equiv  \frac{2mAE}{\pi^{2}}$.  

Taking into account that
\begin{eqnarray}
\label{accior}
I_{2} & = & \sqrt{\mu \epsilon - \mu^{2} I_{1}^{2}} \; ,\\
\omega_{2} & = & \frac {2I_{2}}{\mu} \; ,\\
\left| \partial_{\mu}{h}-\partial_{\mu}{h^{\prime}} \right| & = &
2 \left |  I_{1}^{2} - I_{1}^{\prime 2} \right | 
\end{eqnarray}
and setting $\hbar=1$, we obtain, after replacing in Eq.(~\ref{smooth})
\begin{equation}
\label{suacua}
\left < \rho_{c}(\epsilon,\mu) \right > = \frac{\mu}{4}
\int_{I_{1}^{\prime}=0}^{I_{1}^{\prime}=\sqrt{\epsilon / \mu}} 
\frac{dI_{1}^{\prime}} {\sqrt{\epsilon-\mu I_{1}^{\prime 2}}} 
\left [ \int_{I_{1}=I_{1}^{\prime}}^{I_{1}=\sqrt{\epsilon / \mu}} 
\frac{ (I_{1}^{2} - I_{1}^{\prime 2}) dI_{1}}
{\sqrt{\epsilon-\mu I_{1}^{2}}}
+\int_{I_{1}=0}^{I_{1}=I_{1}^{\prime }} 
\frac{ (I_{1}^{\prime 2} - I_{1}^{2}) dI_{1}}
{\sqrt{\epsilon-\mu I_{1}^{2}}} \right ] \; ,
\end{equation}
that after elementary integrations leads to
\begin{equation}
\label{smoothr}
\left < \rho_{c}(\epsilon,\mu) \right > = \frac{1}{4} \frac{\epsilon}{\mu} \; .
\end{equation}

\begin{figure}
\begin{center}
\leavevmode
\epsfysize=10cm
\epsfbox[19 22 589 569]{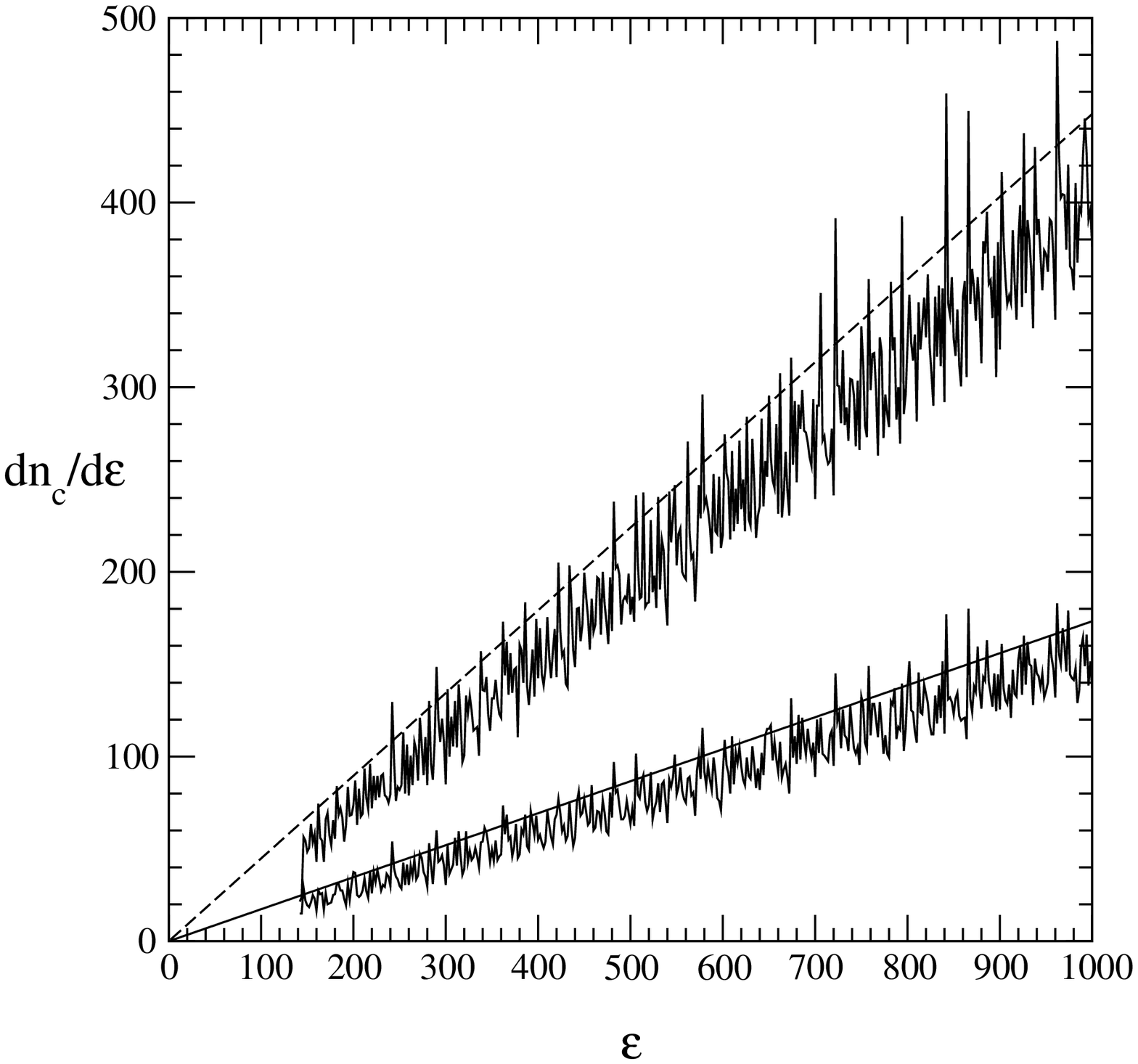}
\end{center}

\caption{Integrated density of crossings $dn_{c}/d\epsilon$
as a function of the energy $\epsilon$ for the rectangular billiard.
The straight lines correspond to the smooth part $\left<dn_{c}/d\epsilon\right>
=\frac{\epsilon}{4} \ln(\frac{\mu_{2}}{\mu_{1}})$ for $\mu_{2}=6$ (dashed line)
and $\mu_{2}=2$ (solid line). In both cases $\mu_{1}=1$.} 
\label{figure1}
\end{figure}

To compare this result with the exact one obtained by quantum calculation, 
we integrate the above density of crossings over the parameter $\mu$. In this way we establish the mean number of crossing in the finite interval $[\mu_{1}, \mu_{2}]$  with energies between $\epsilon$ and $\epsilon +d\epsilon$.
\begin{eqnarray}
\label{pred1}
\left < \frac{dn_{c}}{d\epsilon}(\epsilon,\mu_{1},\mu_{2}) \right > & = &
\int_{\mu_{1}}^{\mu_{2}} \left < \rho_{c}(\epsilon,\mu) \right > d\mu \nonumber\\
  & = & \frac{1}{4} \epsilon \ln{ \left( \frac{\mu_{2}}{\mu_{1}} \right) } \; , 
\end{eqnarray}
which is a linear function of $\epsilon$.
Figure ~\ref{figure1} shows the predictions of the preceding result for 
$\mu_{1}=1$ and
$\mu_{2}=2$ (solid line) and $\mu_{2}=6$ (dashed line) together with the exact calculations. 

\begin{figure}
\begin{center}
\leavevmode
\epsfysize=10cm
\epsfbox[19 22 589 569]{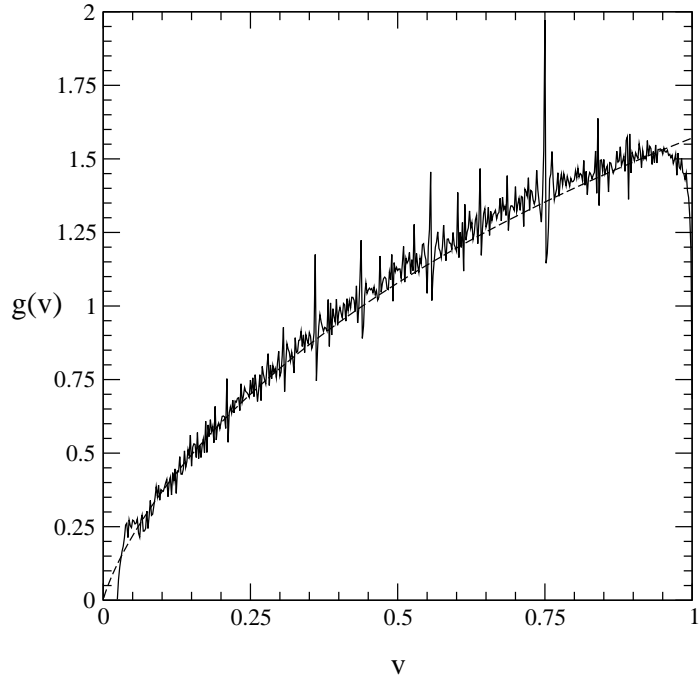}
\end{center}
\caption{Distribution of crossings $g(v)$ according to the relative difference between the slopes $v$ for the rectangular billiard. 
The result predicted for the smooth part of density of crossings corresponds to the dashed line.}
\label{figure2}
\end{figure}

Now we apply Eq.(~\ref{emuv}) to obtain the DCDS.
We define the relative difference of slopes as $v \equiv V/V_{max}$ where 
$V=2 \left| I_{1}^{2}-I_{2}^{2}\right|$ and
$V_{max}$ is the highest value of $V$ for a given value of the energy $\epsilon$
and the parameter $\mu$ (in the present example $V_{max}=2 \epsilon/\mu$).
By changing variables in Eq.(~\ref{suacua}) 
$(I_{1},I_{1}^{\prime}) \rightarrow (z=I_{1} \sqrt{\mu/\epsilon}, v)$ and leaving
out the integration over $v$, we obtain
\begin{equation}
\label{jointca}
\left<\rho_{c}(\epsilon,\mu,v)\right>=\frac{1}{4}
\frac{\epsilon}{\mu} \; g(v) \; ,
\end{equation}
where
\begin{equation}
\label{g1}
g(v)= v \int_{z=\sqrt{v}}^{z=1} \frac{dz}{\sqrt{(1-z^2)(1-(z^2-v))(z^2-v)}} \; .
\end{equation}
Therefore, the joint distribution Eq.(~\ref{jointca}) results factorizable as
$\left<\rho_{c}(\epsilon,\mu)\right> g(v)$. This means that the distribution
of crossings according to the relative difference between the slopes of the levels is
a global property of the system. It holds for any region in the plane $\epsilon-\mu$. Equation (~\ref{g1}) is normalized to one and it gives the fraction of crossings whose relative difference of slopes is in the interval
$[v,v+dv]$ and it is independent on the values of $\epsilon$ and $\mu$. For example, the distribution of the number of crossings in the intervals $[0,\epsilon]$ and $[1,\mu]$ according to the difference of the slopes is
\begin{equation}
\left< \frac{dn_{c}}{dv}\right>= \frac{\epsilon^{2}\ln{\mu}}{8} g(v) \;.
\end{equation}
That is, if we perform an histogram according to the relative difference 
between the slopes for all the crossings in the intervals $[0,\epsilon]$ and $[1,\mu]$, the smooth part of such histogram will be given by Eq.(~\ref{g1}).
Figure ~\ref{figure2} shows the distribution Eq.(~\ref{g1}) 
together with the histogram resulting from the exact quantum calculation computing the relative difference between the slopes of the crossings that occur in the region $\epsilon<3000$ and $1\leq \mu \leq 2$ ($\sim 720000$ crossings).         

\subsection{Aharonov-Bohm cylindrical billiards }

\begin{figure}

\begin{center}
\leavevmode
\epsfysize=10cm
\epsfbox[19 22 589 569]{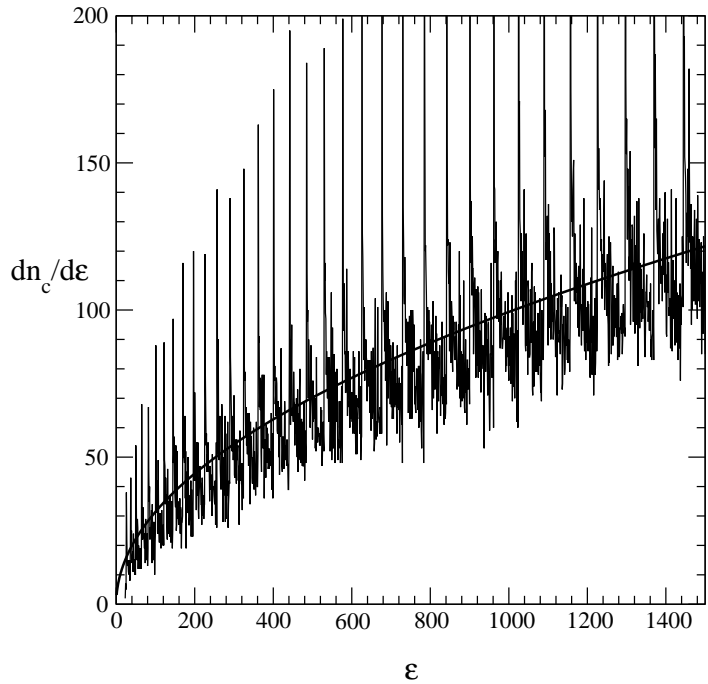}
\end{center}
\caption {Integrated density of crossings $dn_{c}/d\epsilon$
as a function of energy $\epsilon$ for the Aharonov-Bohm cylindrical billiard
with $\gamma=\frac{4}{\pi^2}$.
The solid smooth line corresponds to the smooth part $\left<dn_{c}/d\epsilon\right>
=2\sqrt{\frac{\epsilon}{\gamma}}$. } 
\label{figure3}
\end{figure}

As a second example, we consider the Aharonov-Bohm cylindrical billiard. That 
is a spinless particle confined in a two dimensional cylindrical shell of height $a$ and radius $r$ axially threaded by a confined magnetic flux $\phi$. 
The hamiltonian is (taking $I_{1}$ non negative)
\begin{equation}
H(I_{1},I_{2})= \frac{(I_{1} \pm \frac{q\Phi}{2\pi c})^{2}}{2mr^{2}}
+\frac{\pi^{2} I_{2}^{2} }{2ma^{2}} \; ,
\end{equation}
that can be rewritten as
\begin{equation}
h = \gamma (I_{1} \pm \phi)^{2} + I_{2}^{2} \; ,
\end{equation}
where $\gamma \equiv\frac{a^2}{\pi^{2}r^{2}}$, $\phi\equiv\frac{q\Phi}{2\pi c}$ and
$h\equiv \frac{2ma^{2}H}{\pi^2}$. As in the first example we define the 
energy $\epsilon \equiv \frac{2ma^{2}E}{\pi^2}$.
Now, we will consider the crossings as a function of the normalized magnetic flux $\phi$. That is we set $\mu=\phi$.
Taking into account that
\begin{eqnarray}
I_{2} & = & \sqrt{ \epsilon - \gamma (I_{1} \pm \mu)^{2}} \; , \\
\omega_{2} & = & 2I_{2} \; ,\\
\left| \partial_{\mu}{h}-\partial_{\mu}{h^{\prime}} \right| & = &
2 \gamma \left |  I_{1} \pm I_{1}^{\prime} \right | \; , 
\end{eqnarray}
and setting $\hbar=1$, we obtain:
\begin{eqnarray}
\label{flux}
\left < \rho_{c}(\epsilon,\mu) \right >& =& \gamma
\int_{I_{1}^{\prime}=0}^{I_{1}^{\prime}=\sqrt{\epsilon / \gamma}} 
\frac{dI_{1}^{\prime}} {\sqrt{\epsilon-\gamma I_{1}^{\prime 2}}} 
\left [  \int_{I_{1}=I_{1}^{\prime}}^{I_{1}=\sqrt{\epsilon / \gamma}} 
\frac{ (I_{1} - I_{1}^{\prime }) dI_{1}}
{2 \sqrt{\epsilon-\gamma I_{1}^{2}}} \right. \nonumber \\
& & \left.+\int_{I_{1}=0}^{I_{1}= I_{1}^{\prime}} 
\frac{ (I_{1}^{\prime } - I_{1}) dI_{1}}
{2 \sqrt{\epsilon-\gamma I_{1}^{2}}}+
\int_{I_{1}=0}^{I_{1}=\sqrt{\epsilon / \gamma}} 
\frac{ (I_{1}+I_{1}^{\prime})dI_{1}} {2 \sqrt{\epsilon-\gamma I_{1}^{2}}} 
\right] \;.
\end{eqnarray}
To derive Eq.(\ref{flux}) it is necessary to take into account in 
Eq.(\ref{rho}), the effect of the breaking of the time reversal invariance (see Appendix).
After performing the integrations, Eq.(\ref{flux}) gives
\begin{equation}
\label{cilcru}
\left < \rho_{c}(\epsilon,\mu) \right > = 
\frac{2\sqrt{\epsilon}}{\sqrt{\gamma}} \; .
\end{equation}
This smooth density of crossings is independent on the parameter $\mu$ (the flux). Therefore the number of crossing between $\epsilon$ and $\epsilon+d\epsilon$ 
per unit of flux is given by Eq.(~\ref{cilcru}).
Figure ~\ref{figure3}  shows $\left < \rho_{c}(\epsilon,\mu) \right >$
given by Eq.(~\ref{cilcru}) and the 
exact quantum calculation for a system with $\gamma= \frac{4}{\pi^{2}}$.    

Starting from Eq.(~\ref{flux}), it is not difficult to establish the DCDS. Let distinguish the crossings according to the relative sign between the slopes. We label by a plus sign $(+)$ (minus sign, $(-)$) the crossings 
between levels with equal (different) sign of their slopes.
We can discriminate in expression Eq.(~\ref{flux}) the contributions from
both kind of crossings. The first and the second integral in the
square bracket correspond to crossings with the same sign of the slopes
while the third integral corresponds to crossings with different sign.
Thus we find
\begin{eqnarray}
\label{fracci}     
\left < \rho_{c}(\epsilon,\mu) \right > & = & 
\left < \rho_{c}(\epsilon,\mu)^{+} \right >   
+\left < \rho_{c}(\epsilon,\mu)^{-} \right >  \; , \nonumber \\
\left < \rho_{c}(\epsilon,\mu)^{+} \right > & = & 
\sqrt{\frac{\epsilon}{\gamma}} \left( 2-\frac{\pi}{2} \right) \; ,\nonumber \\ 
\left < \rho_{c}(\epsilon,\mu)^{-} \right > & = & 
\sqrt{\frac{\epsilon}{\gamma}} \frac{\pi}{2} \; .  
\end{eqnarray}
Although the smooth part of the density of crossings depends on the 
energy $\epsilon$ and on the shape parameter $\gamma$, the fractions of each
kind of crossings are the same for all cylinders and they are
independent on the energies.

To establish the DCDS we proceed as follows. We define the relative jump of the slopes as:
\begin{equation}
v= \sqrt{\frac{\gamma}{\epsilon}} \left| I_{1}-I_{1}^{\prime}\right| \; ,
\end{equation}
for crossings with same sign of the slopes and
\begin{equation}
v=\sqrt{\frac{\gamma}{\epsilon}} (I_{1}+I_{1}^{\prime}) \; ,
\end{equation}
for crossings with different sign of the slopes. In this way  
$v$ results $0 \leq v \leq 1$  for crossings labeled $(+)$
while it is $0 \leq v \leq 2$ for crossings $(-)$.

Thus by changing variables in
Eq.(~\ref{flux}) and leaving out the integration over $v$ we obtain
\begin{equation}
\label{dislope}
\left < \rho_{c}(\epsilon,\mu,v) \right > = \left\{ \begin{array}{ll}
2\sqrt{\frac{\epsilon}{\gamma}}
\left(g_{I}(v)^{+} + g_{I}(v)^{-}\right)     & \mbox{if $v \leq 1$} \\
2 \sqrt{\frac{\epsilon}{\gamma}} \; g_{II}(v)^{-} & \mbox{if $2 \geq v > 1$} \;,
\end{array}
\right.
\end{equation}
where
\begin{eqnarray}
\label{gvv}
g_{I}(v)^{+} & = & \frac{1}{2} v \int_{v}^{1} 
\frac{dz}{\sqrt{1-(z-v)^2}\sqrt{1-z^2}} \nonumber \; , \\
g_{I}(v)^{-} & = & \frac{1}{4} v \int_{0}^{v} 
\frac{dz}{\sqrt{1-(v-z)^2}\sqrt{1-z^2}} \nonumber \; ,\\
g_{II}(v)^{-} & = & \frac{1}{4} v \int_{v-1}^{1} 
\frac{dz}{\sqrt{1-(v-z)^2}\sqrt{1-z^2}}  \;.
\end{eqnarray}
Equation (\ref{dislope}) gives the mean number of crossings per unit of flux
that occur in the interval of energy
$[\epsilon,\epsilon+d\epsilon]$ and such that the relative difference
between the slopes lies in the interval $[v,v+dv]$ per unit of flux.
For $v \leq 1$ there are two contibutions. The first integral corresponds 
to crossings with the same sign of the slopes and the second one corresponds
to crossings between levels with different sign of the slopes. For $v > 1$, only
crossings with different sign of the slopes can occur. Let us remark that
the joint distribution  $\left < \rho_{c}(\epsilon,\mu,v) \right >$
results factorizable (as in the rectangular billiard)
\begin{equation}
\left < \rho_{c}(\epsilon,\mu,v) \right >=
\left < \rho_{c}(\epsilon,\mu) \right > \; g(v) \; ,
\end{equation}
with
\begin{equation}
\label{gv}
g(v) = \left\{ \begin{array}{ll}
g_{I}(v)^{+} + g_{I}(v)^{-}   & \mbox{if $v \leq 1$} \; ,\\
g_{II}(v)^{-}      & \mbox{if $1 < v \leq 2$} \; .
\end{array}
\right.
\end{equation}
As in the rectangular billiard, distribution Eq.(~\ref{gv}) holds
irrespective of the values of the flux and the energy.
Therefore, the distribution of crossing below an energy
$\epsilon$ according to the value of $v$ will be given by
\begin{eqnarray}
\left< \frac{dn_{c}(\epsilon,\mu,v)}{dv}\right> & = & g(v) \int_{0}^{\epsilon}
 \left < \rho_{c}(\epsilon^{\prime},\mu) \right > d\epsilon^{\prime} 
\nonumber\\
             & = & \frac{4}{3} \frac{(\epsilon)^{3/2}}{ \sqrt{\gamma}} \; g(v) \; .
\end{eqnarray}

\begin{figure}
\begin{center}
\leavevmode
\epsfysize=10cm
\epsfbox[19 22 589 569]{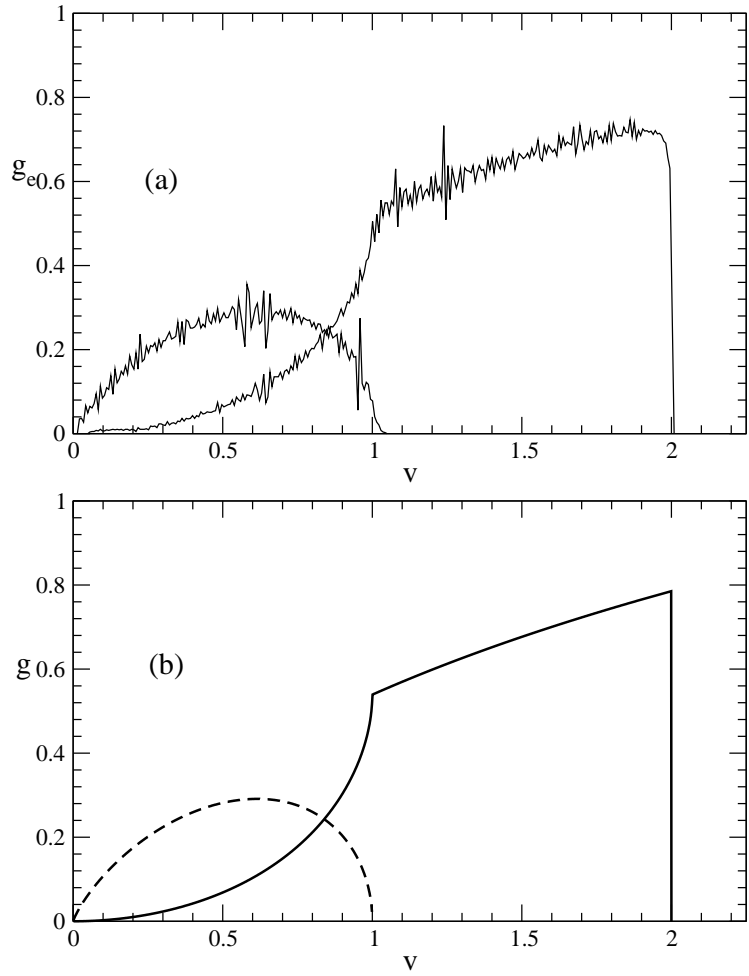}
\end{center}
\caption{Distribution of crossings $g(v)$ according to the relative difference between the slopes $v$. 
a) Exact quantum results. b) Results predicted for the smooth part of the
density of crossings.The dashed line curve corresponds to crossings
between levels with equal sign of their slopes, while the solid line corresponds to crossings between levels with different sign.}
\label{figure4}
\end{figure}

Figure ~\ref{figure4} shows $g(v)$ predicted by Eq.(~\ref{gvv}) together with
the exact quantum results obtained computing the crossings for $\epsilon< 1400$ between $0 \leq \mu <1$ ($n_{c}=105158$ crossings). We have discriminated the crossings according to the same sign ($g_{I}(v)^{+}$, dashed line) or different sign ($g_{I}^{-}+g_{II}^{-}$, solid line) between the slopes of the levels. We remark that there are
$n_{c}^{+}=22266$ crossings $(+)$ and  $n_{c}^{-}=82892$ crossings $(-)$.These results are consistent with Eq.(~\ref{fracci}) that implies
\begin{eqnarray}
\frac{n_{c}^{+}}{n_{c}} &=& 2-\frac{\pi}{2}  \; ,\nonumber\\  
\frac{n_{c}^{-}}{n_{c}} &=& \frac{\pi}{2} \; .
\end{eqnarray}

\section{The oscillating contributions}
\label{oscpart}
In this section we calculate the oscillating contributions to the density of
crossings Eq.(~\ref{osmain}).  
They are determined by the periodic orbits that satisfy condition 
Eq.(~\ref{perio}). 

\subsection{Rectangular billiards}
\label{oscrec}

Given the hamiltonian Eq.(~\ref{hamilr}) the periodic orbits are classified
according to the topology $(m_{1},m_{2})$ being $m_{1} (m_{2})$ the number of bounces on a side of length $b(a)$ before the periodic orbits are closed.

\begin{figure}
\begin{center}
\leavevmode
\epsfysize=8.2cm
\epsfbox[19 22 589 569]{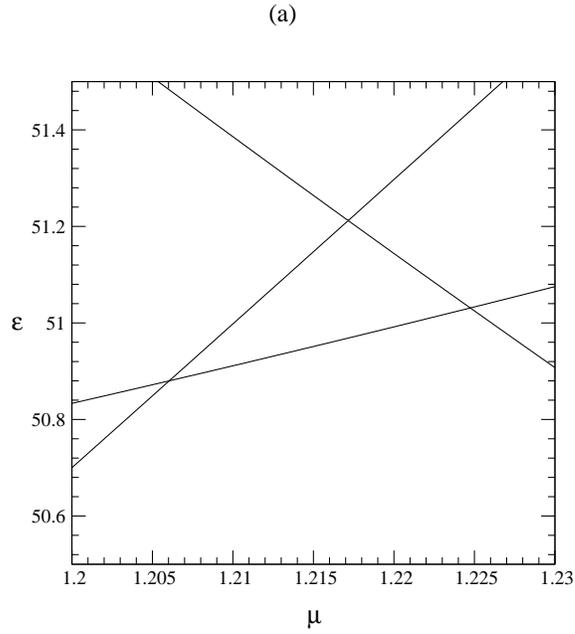}
\end{center}
\vspace{1.3cm}
\begin{center}
\leavevmode
\epsfysize=10cm
\epsfbox[19 22 589 569]{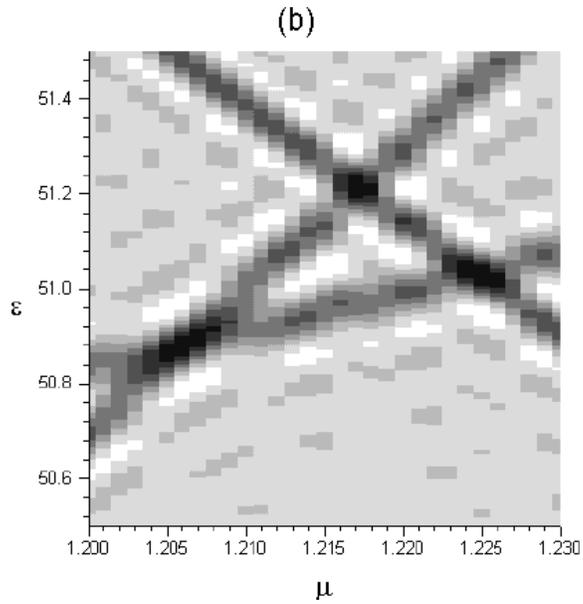}
\end{center}

\caption{a) Detail of three exact quantum eigenergies for the rectangular billiard as a function of $\mu$  where three crossings can be observed. b) 
Density plot of the oscillating part $\rho_{{c}_{osc1}}$ (see ~\ref{oscrec} in
the text) of the density of crossings in the same region of the plane
$\epsilon-\mu$ .} 
\label{figure5}
\end{figure}

Taking into account Eq.(~\ref{hamilr}) and the requirement Eq.(~\ref{perio}),
we have:
\begin{equation}
\frac{I_{1}^{o} \mu^{2}}{I_{2}^{o}} =  \frac{m_{1}}{m_{2}} \; , \\
\end{equation}
and
\begin{eqnarray}
\label{acpere}
I_{1}^{o}(m_{1},m_{2}) & = & \left( \frac{\epsilon}
{\frac{m_{1}^{2}}{\mu}+m_{2}^{2} \mu} \right)^{1/2} 
\frac{m_{1}}{\mu} \; , \nonumber \\
I_{2}^{o}(m_{1},m_{2}) & = & \left( \frac{\epsilon} 
{\frac{m_{1}^{2}}{\mu}+m_{2}^{2} \mu} \right)^{1/2} m_{2} \mu \; .
\end{eqnarray}
Therefore from Eqs.(~\ref{action}) and (~\ref{theta}) follows:
\begin{eqnarray}
\label{srec}
S(m_{1},m_{2}) & = & 2\pi \sqrt{\left(\frac{m_{1}^{2}}{\mu}+m_{2}^{2} \mu\right) \; \epsilon} \; , \\
\theta(m_{1},m_{2}) & = & -\frac{\pi}{4} \; .
\end{eqnarray}
On the other hand, using expression Eq.(~\ref{accior}), we have
\begin{equation}
\label{derirec}
\left| \frac{d^{2}I_{2}}{dI_{1}^{2}} \right|  =  \frac{\mu^{3} 
\epsilon}{I_{2}^{3}} \; . 
\end{equation}
We replace Eqs.(~\ref{acpere}),(~\ref{srec}) and 
(~\ref{derirec}) in Eq.(~\ref{osmain}) and performing elemental integrations
we obtain the oscillating contributions:
\begin{eqnarray}
\rho_{c}(\epsilon,\mu)_{osc1} &=& \sum_{m_{1},m_{2} \neq(0,0)}
\left(\frac{\epsilon}{\mu}\right)^{3/4} \frac{1}{(m_{1}^{2}+m_{2}^{2} \mu^{2})^{1/4}}
\left[ \frac{(4 m_{1}m_{2}\mu+\pi m_{2}^{2}\mu^{2}-\pi m_{1}^{2})}
{8(m_{1}^{2}+m_{2}^{2} \mu^{2})}\right. \nonumber\\
& & \left. +\frac{(m_{1}^2-m_{2}^{2} \mu^{2})}{2(m_{1}^{2}+m_{2}^{2}\mu^{2})}
\arcsin{\left( \frac{m_{1}}{\sqrt{m_{1}^2+m_{2}^{2}\mu^{2}}}\right) } 
\right] \nonumber \\
&  & \times \cos{\left( 2\pi  \sqrt{(\frac{m_{1}^{2}}{\mu}+m_{2}^{2} \mu) \epsilon}
-\frac{\pi}{4} \right) } \; .
\end{eqnarray}
On the other hand, replacing Eqs.(~\ref{acpere}),(~\ref{srec}) and 
(~\ref{derirec}) in Eq.(~\ref{gral}) we obtain
\begin{eqnarray}
\rho_{c}(\epsilon,\mu)_{osc2}&=& \sum_{m_{1},m_{2} \neq (0,0)} 
\sum_{m_{1},m_{2} \neq (0,0)} \left(\frac{\epsilon}{\mu}\right)^{1/2} 
\frac{1}{(m_{1}^2+m_{2}^2\mu^{2})^{1/4} 
(m_{1}^{\prime 2}+m_{2}^{\prime 2} \mu^{2})^{1/4}} \nonumber \\
& &\times\left| \frac{m_{1}^{2}}{(m_{1}^2+m_{2}^2 \mu^2)} -
\frac{m_{1}^{\prime 2}}{(m_{1}^{\prime 2}+m_{2}^{\prime 2} \mu^{2})}
\right| \nonumber \\
&  & \times \cos{\left( 2\pi\sqrt{(\frac{m_{1}^{2}}{\mu}+m_{2}^{2} \mu) 
\epsilon}
-\frac{\pi}{4} \right)} \cos{\left( 2\pi  \sqrt{(\frac{m_{1}^{\prime 2}}
{\mu}+m_{2}^{\prime 2} \mu) \epsilon}
-\frac{\pi}{4} \right) } \; .  
\end{eqnarray} 
Unlike the smooth part Eq.(~\ref{smoothr}), both contributions involve 
information about the individual crossings. 
The calculation of $\rho_{c}(\epsilon,\mu)_{osc2}$ implies a
quadruple sum over $m's$ while $\rho_{c}(\epsilon,\mu)_{osc1}$   
is a double sum. However if we consider only   
$\rho_{c}(\epsilon,\mu)_{osc1}$ we have a satisfactory resolution of the individual crossings in the plane $\epsilon-\mu$ as can be seen in 
Fig.~\ref{figure5}. This figure shows a region in the plane $\epsilon-\mu$
where there are three crossings. In Fig.~\ref{figure5}(a) 
the exact quantum levels are plotted. Figure ~\ref{figure5}(b) shows the oscillating part Eq.(~\ref{osmain}) considering $m_{1}$ and $m_{2}$ up to 150.  

\subsection{Aharonov-Bohm cylindrical billiards}
\label{oscil}
In this example, the periodic orbits are labeled by the number of 
rotations around the axis of the cylinder ($m_{1}$) and the number of bounces on the base of the cylindrical surface ($m_{2}$) before the periodic orbits are closed.
Taking into account that 
\begin{equation}
\frac{I_{1}^{o} \gamma}{I_{2}^{o}} = \frac{m_{1}}{m_{2}} \; ,
\end{equation}
and
\begin{eqnarray}
\label{acperec}
I_{1}^{o}(m_{1},m_{2}) & = & \left( \frac{\epsilon}
{\gamma(m_{1}^{2}+\gamma m_{2}^{2}) } \right)^{1/2} m_{1} \; ,\nonumber \\
I_{2}^{o}(m_{1},m_{2}) & = & \left( \frac{\gamma \epsilon} 
{(m_{1}^{2}+\gamma m_{2}^{2} } \right)^{1/2} m_{2} \; , \nonumber \\ 
S(m_{1},m_{2}) &=& 2\pi \sqrt{\frac{\epsilon}{\gamma}\left(m_{1}^{2}+
\gamma m_{2}^{2}\right)} \; ,\nonumber\\
\frac{d^{2}I_{2}}{dI_{1}^{2}}&=& \frac{\gamma \epsilon}{I_{2}^{3}} \; , 
\end{eqnarray}
We obtain (for details see the Appendix):
\begin{eqnarray}
\label{osc1ci}
\rho_{c}(\epsilon,\mu)_{osc1} &=& \sum_{m_{1},m_{2} \neq(0,0)}
\Delta_{m_{1}} \left [ \frac{\epsilon}{\gamma(m_{1}^2+\gamma m_{2}^2)}
\right ]^{1/4} \left[ \frac{m_{1}}{\sqrt{m_{1}^2+\gamma m_{2}^2}}
\arcsin{\left(\frac{m_{1}}{\sqrt{m_{1}^2+\gamma m_{2}^2}}
 \right)} \right. \nonumber \\
& & \left.+\frac{\sqrt{\gamma} m_{2}}{\sqrt{m_{1}^2+\gamma m_{2}^2}} \right ] \nonumber\\
 & & \times \cos{\left( 2\pi \sqrt{\frac{\epsilon}{\gamma}(m_{1}^{2}+
\gamma m_{2}^{2})}-\frac{\pi}{4} \right)} \cos{(2\pi m_{1} \mu)} \; ,
\end{eqnarray}
where $\Delta_{m_{1}}=2$ if $m_{1}\neq 0$ and $\Delta_{0}=1$, and 
\begin{eqnarray}
\label{osc2ci}
\rho_{c}(\epsilon,\mu)_{osc2} &=& \sum_{m_{1},m_{2} \neq (0,0)} 
\sum_{m_{1}^{\prime},m_{2}^{\prime} \neq (0,0)}
\left|\frac{m_{1}}{(m_{1}^2+\gamma m_{2}^2)}
-\frac{m_{1}^{\prime}}{(m_{1}^{\prime 2}+\gamma m_{2}^{\prime 2})}
 \right| \nonumber \\
& & \times \left[ 
\cos{\left(2\pi \left(\sqrt{\frac{\epsilon}{\gamma}(m_{1}^{2}+
\gamma m_{2}^{2})} - \sqrt{\frac{\epsilon}{\gamma}(m_{1}^{\prime 2}+
\gamma m_{2}^{\prime 2})}\right) \right)} \cos{\left(2\pi(m_{1}-m_{1}^{\prime})\mu\right)} \right. 
\nonumber \\
& & \left.+ \cos{\left(2\pi \left(\sqrt{\frac{\epsilon}{\gamma}(m_{1}^{2}+
\gamma m_{2}^{2})}+ \sqrt{\frac{\epsilon}{\gamma}(m_{1}^{\prime 2}+
\gamma m_{2}^{\prime 2})}-1/4 \right)\right)}
\cos{\left(2\pi(m_{1}+m_{1}^{\prime})\mu\right)} 
\right] \nonumber \\
& & +\left[ \frac{m_{1}}{(m_{1}^2+\gamma m_{2}^2)}
+\frac{m_{1}^{\prime}}{(m_{1}^{\prime 2}+
\gamma m_{2}^{\prime 2})}\right] \nonumber \\
& & \times \left[ \cos{\left(2\pi \left(\sqrt{\frac{\epsilon}{\gamma}
(m_{1}^{2}+\gamma m_{2}^{2})}+ \sqrt{\frac{\epsilon}{\gamma}
(m_{1}^{\prime 2}+
\gamma m_{2}^{\prime 2})}-1/4\right)\right)}\cos{\left(2\pi(m_{1}-m_{1}^{\prime})\mu \right)} 
\right.\nonumber \\
& &\left.+  \cos{\left(2\pi\left(\sqrt{\frac{\epsilon}{\gamma}(m_{1}^{2}+
\gamma m_{2}^{2})}- \sqrt{\frac{\epsilon}{\gamma}(m_{1}^{\prime 2}+
\gamma m_{2}^{\prime 2})}\right)\right)}\cos{\left(2\pi(m_{1}+m_{1}^{\prime})\mu \right)} \right] \; .
\end{eqnarray}

\begin{figure}
\begin{center}
\leavevmode
\epsfysize=8.2cm
\epsfbox[19 22 589 569]{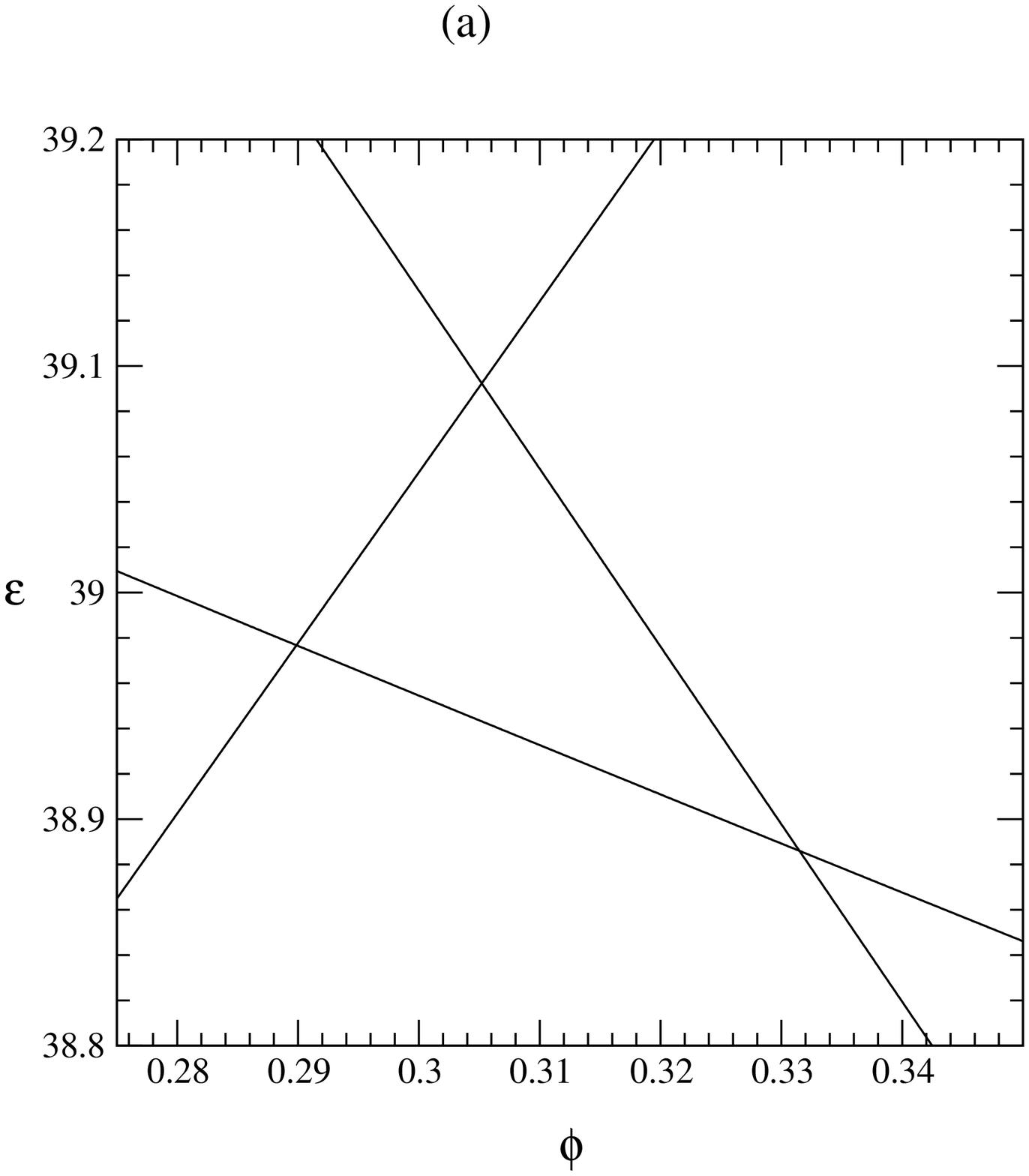}
\end{center}
\vspace{1.3cm}
\begin{center}
\leavevmode
\epsfysize=10cm
\epsfbox[19 22 589 569]{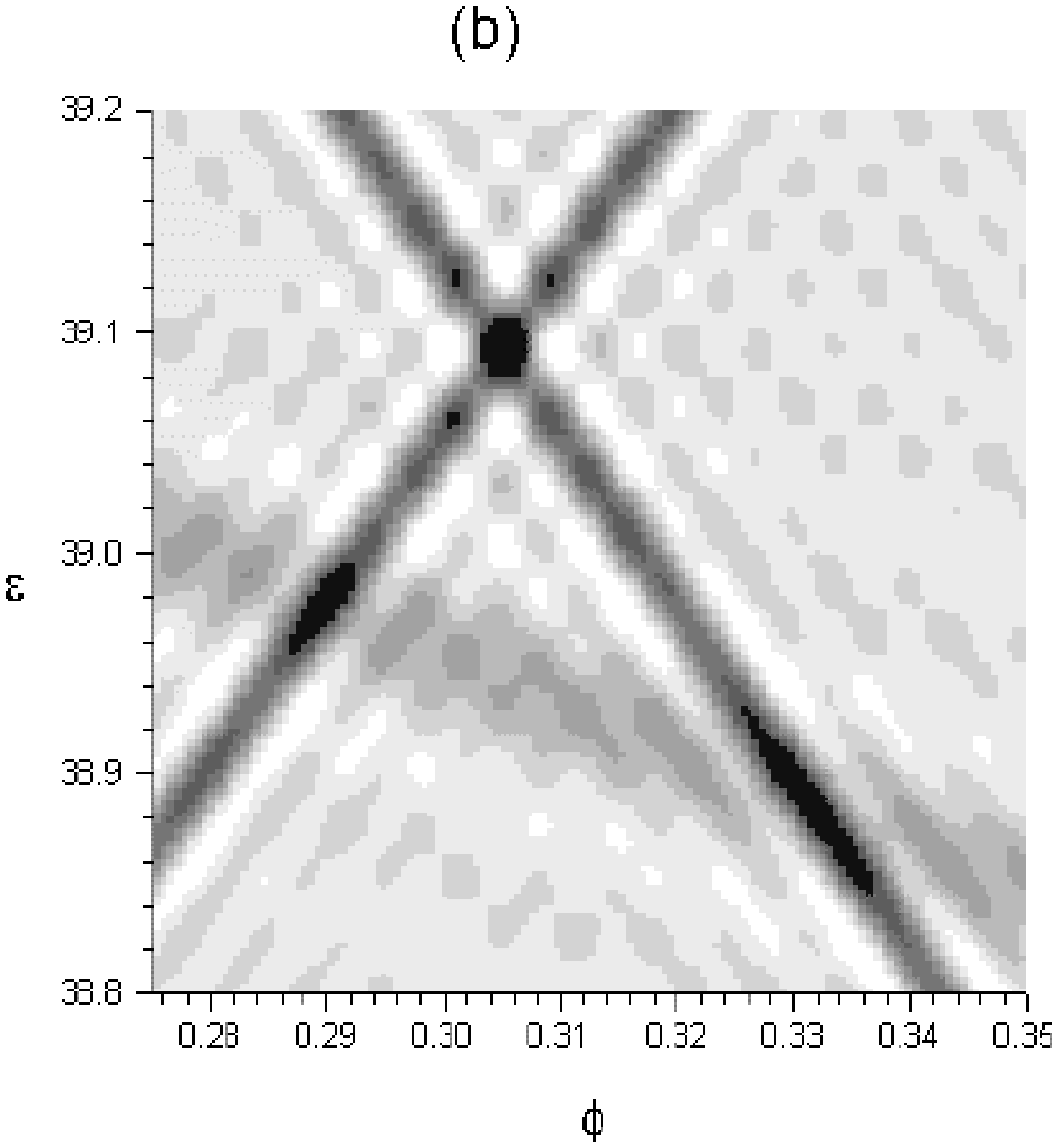}
\end{center}

\caption{a) Detail of three exact quantum eigenergies for the Aharonov- Bohm cylindrical billiard as a function of the flux $\phi$ where three crossings can be observed. b) Density plot of the oscillating part $\rho_{{c}_{osc1}}$ 
(see ~\ref{oscil} in the text) of the density of crossings in the same region of the plane $\epsilon-\phi$.} 
\label{figure6}
\end{figure}

As in the example of the previous subsection, 
$\rho_{c}(\epsilon,\mu)_{osc1}$ Eq.(~\ref{osc1ci}), gives enough information
to determine individual crossings, as can be seen in Fig. ~\ref{figure6}.
In Fig. ~\ref{figure6}(a) is depicted a region of the plane $\epsilon-\mu$
where the exact quantum levels show three crossings. 
Figure ~\ref{figure6}(b) shows the oscillating contribution
given by Eq.(~\ref{osc1ci}) taking into account $m_{1}$ and $m_{2}$ up to
150.

\begin{figure}
\begin{center}
\leavevmode
\epsfysize=10cm
\epsfbox[19 22 589 569]{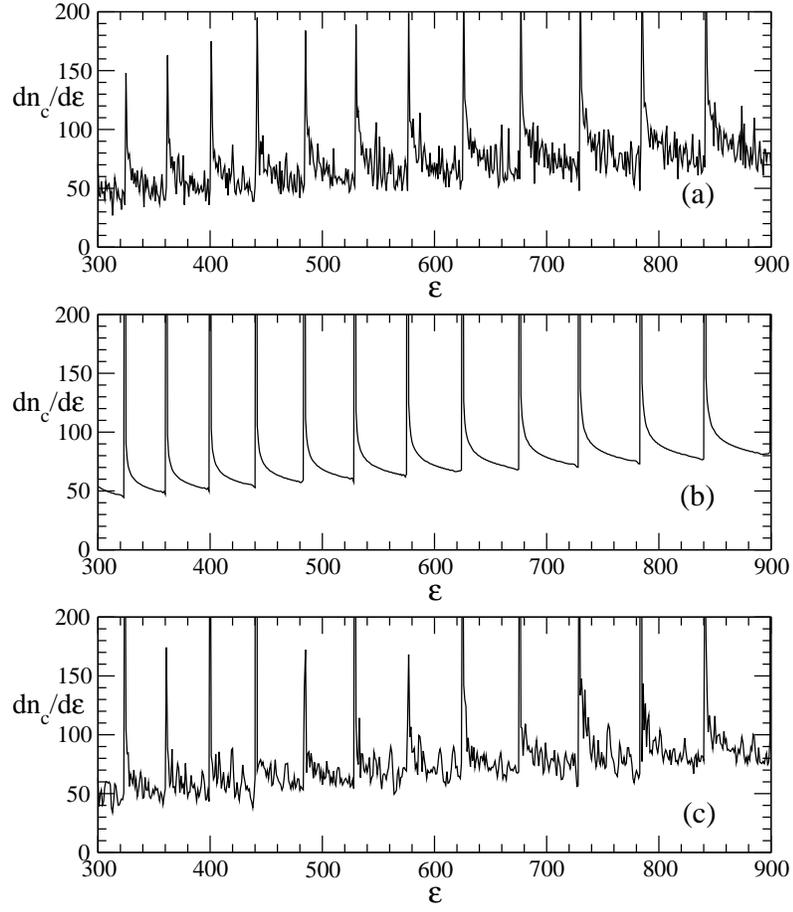}
\end{center}
\caption{a)Integrated density of crossings $dn_{c}/d\epsilon$
as a function of energy $\epsilon$ for the Aharonov-Bohm cylindrical billiard
with $\gamma=\frac{4}{\pi^2}$.
b) The same density predicted by the sum of the smooth part $\left<\rho_{c}\right>$ and the contribution originated by $\rho_{{c}_{osc1}}$.
c) The same density predicted by $\left< \rho_{c} \right>+\rho_{{c}_{osc1}}+\rho_{{c}_{osc2}}$
(see ~\ref{oscil} in the text).} 
\label{figure7}
\end{figure}

Unlike the rectangular billiard, in the present case, the dependence on 
the parameter $\mu$ of the oscillating contributions is quite simple.
Therefore Eqs.(~\ref{osc1ci}) and (~\ref{osc2ci}) can be easily integrated to obtain the oscillating part of the number of crossings in the energy interval $[\epsilon, \epsilon+d\epsilon]$ per unity of flux,
\begin{eqnarray}
\frac{dn_{c}}{d\epsilon}& = & \int_{\mu=0}^{\mu=1} \rho_{c}(\epsilon,\mu) 
d\mu \nonumber \\
\label{titi} & = & \int_{\mu=0}^{\mu=1} \left<\rho_{c}\right> d\mu + \int_{\mu=0}^{\mu=1}
\rho_{c}(\epsilon,\mu)_{osc1} d\mu +\int_{\mu=0}^{\mu=1}\rho_{c}(\epsilon,\mu)_{osc2} d\mu \; .
\end{eqnarray}
As we have pointed out in section ~\ref{suave}, the smooth part of the density of crossings depends only on the energy $\epsilon$, so the integrated smooth part (the first integral in the right hand side of Eq.(~\ref{titi}))
is given by Eq.(~\ref{cilcru}). For the oscillating contributions, 
the integration over $\mu$ of the terms in Eq.(~\ref{osc1ci}) vanishes 
unless $m_{1}=0$ (because they are proportional to $\sin(2\pi m_{1})/2\pi m_{1}$). Therefore we obtain:
\begin{equation}
\label{integ1}
\int_{\mu=0}^{\mu=1} \rho_{c}(\epsilon,\mu)_{osc1} d\mu = 
2 \left(\frac{\epsilon}{\gamma}\right)^{1/4} \sum_{m_{2}>0} 
\frac{1}{m_{2}^{1/2}} \cos{(2\pi m_{2}\sqrt{\epsilon}-\frac{\pi}{4})} \; .
\end{equation}
In terms of the classical motion, the contribution Eq.(~\ref{integ1})
is originated by the orbits that having $m_{1}=0$, correspond to an irrotational motion with bounces between the bases of the cylinder.    
Figure ~\ref{figure7}(a) shows the exact quantum calculation of the integrated
density of crossings obtained through an histogram. The sum of the
contribution Eq.(~\ref{integ1}) (up to $m_{2}=500$) and the smooth part Eq.(~\ref{cilcru}) is shown in Fig.~\ref{figure7}(b). We can see 
that the contribution of the oscillating term  Eq.(~\ref{integ1}) originates
the sharp peaks that are present in the exact calculation. By inspection of Eq.(~\ref{integ1}), we expect a sharp peak whenever $\epsilon$ tunes a value such that
\begin{equation}
m_{2}\sqrt{\epsilon}-1/8=l \; ,
\end{equation}
for all $m_{2}$ being $l$ an integer. For $\epsilon >> 1$ this condition leads
to $\epsilon \approx n^2$. Therefore we conclude that the peaks correspond to
the regions of energy where a ``head-rotational band'' state appears. Such states (the first state for a given $n$ ) are the less affected by the change of the flux ($\mu$) (they have the smallest slope because they have angular 
momentum equal to zero) and they contribute
to the density of crossings in a relevant way because they will cross with
all the other states that go through (upward or downward) this region of energy. 
The integration of the terms corresponding to the other oscillating contribution Eq.(~\ref{osc2ci}) also vanishes unless $m_{1}=m_{1}^{\prime}$. Taking into
account these contributions, the integrated density of crossings is shown in fig.~\ref{figure7}(c).

\section{Concluding Remarks}
\label{cr}
In the present paper we have derived a semiclassical expression for the density of degeneracies, $\rho_{c}(E,\mu)$ that occur in quantum integrable systems depending on a single parameter $\mu$. We have applied our results to two specific systems that depend on the parameter in a different functional form, obtaining a quite satisfactory agreement with the exact quantum calculation.
Our results show that the density of crossings is strongly dependent on the derivative of the hamiltonian with respect to the parameter. Therefore, unlike the density of states, even the smooth part of the density of crossings can depend  on the energy $E$ and on the parameter $\mu$ in a diverse way according to the functional form of $H(\mu)$. In particular, the smooth part of $\rho_{c}$ for the rectangular billiard, where $\mu$ is the ratio between its sides, results $\sim E/\mu$.
Taking into account that the origin of the crossings between eigenenergies of integrable billiards as a function of a shape parameter has the same and very simple interpretation in terms of geometrical arguments \cite{traiber}, we expect that the dependence $\sim E/\mu$ of the smooth part of the density of crossings, holds for any integrable billiard (like the elliptic billiard where $\mu$ is the ratio between the axes of the elliptical box or the annular billiard where $\mu$ is the ratio between the outer and inner radii).
On the other hand, for the Bunimovich stadium \cite{vergi}, there are numerical evidences that the smooth part of the integrated density of narrowly avoided crossings (that is the number of narrowly avoided crossings per unit of $E$
for a given interval of $\mu$, being $\mu$ the shape parameter), shows a linear dependence on the energy $E$.  
This suggests that the integrated distribution of narrowly avoided crossings when $\mu$ is a shape parameter would have the same functional dependence on $E$ (linear) for other irregular billiards.

For the Aharonov-Bohm cylindrical billiard, the smooth part of the density of crossings is $\sim \sqrt{E}$ without dependence on $\mu=\phi$ (the magnetic flux). Such functional form should hold for other Aharonov-Bohm integrable billiards
and this is the case for the Aharonov-Bohm annular billiard \cite{majale}. 
Moreover in Ref.\cite{majale} it is shown, by numerical computation, that the distribution of spacing
in flux between crossings that a given level has, is Poissonian. The present
work provides the framework to study properties of the crossing spacing distribution, employing the two point correlation function for the density of degeneracies.

\appendix

\section*{}

The present appendix is devoted to show how to handle the sum Eq.(~\ref{rho})
for the Aharonov-Bohm cylindrical billiard.
This system having one degree of freedom of rotation, it has not
Kramer's degeneracies. That is, conjugate states by time reversal
transformation are non degenerated.

At first, let us consider the billiard without flux. The eigenenergies
are
\begin{equation}
h(n_{1},n_{2})= \gamma n_{1}^2 + (n_{2}+1)^2 \; ,
\end{equation}
where $n_{1}$ is the quantum number related to the rotation around the
axis of the cylinder $(n_{1}=0,\pm1,\pm2,...)$ and $n_{2}$ is the quantum 
number related to libration
motion parallel to axis $(n_{2}=0,1,2,3)$. The system has Kramers's degeneracy.
A quantum eigenstate $(n_{1},n_{2})$ and its conjugate $(-n_{1},n_{2})$ have
the same eigenenergy if $n_{1} \neq 0$.
To take into account these degeneracies in the sum Eq.(~\ref{rho}) over 
$n_{1},n_{1}^{\prime} \geq 0 $ we must include a factor 
$e(n_{1},n_{1}^{\prime})$ in each term such that
\begin{equation}
e(n_{1},n_{1}^{\prime}) = \left\{ \begin{array}{ll}
4   & \mbox{if $n_{1} \neq 0$ and $n_{1}^{\prime} \neq 0$} \\
2      & \mbox{if $(n_{1} \neq 0$ and $n_{1}^{\prime}=0)$ or 
              $(n_{1} \neq 0$ and $n_{1}^{\prime} \neq 0$)} \\ 
0   & \mbox{if $n_{1} =  0$ and $n_{1}^{\prime} =0$} \; .
\end{array}
\right.
\end{equation}

When the cylinder is threaded by a magnetic flux $\phi$, the 
eigenenergies of a conjugate pairs of eigenstates are
\begin{eqnarray}
h(n_{1},n_{2},\phi) &=&  \gamma (n_{1}-\phi)^2 + (n_{2}+1)^2  \nonumber \; , \\ 
h(-n_{1},n_{2},\phi) &=& \gamma (-n_{1}-\phi)^2 + (n_{2}+1)^2  \nonumber \; ,\\ 
                & = & \gamma (n_{1}+\phi)^2 + (n_{2}+1)^2 \; .
\end{eqnarray}
Therefore, conjugated pairs of eigenstates are no longer degenerated.
In such a situation, to preserve the sum Eq.(~\ref{rho}) over 
$n_{1},n_{1}^{\prime} \geq 0 $, we must distinguish two kinds of states
according to the dependence on the flux that their eigenenergies have. 
Moreover, Eq.(~\ref{para}) which determines the values of the parameter
for which the crossings occur, changes according to the same dependence.
Thus, Eq.(~\ref{rho}) for the cylindrical billiard means
\begin{eqnarray} 
\label{rhocil}
\rho_{c}(\epsilon,\phi)& = &\frac{1}{2}\sum_{\vec{n}} 
\sum_{\vec{n}^{\prime} } 
\delta{\left(\epsilon-h(n_{1}+\phi,n_{2})\right)} 
\delta{\left(\phi-L_{1}(n_{1},n_{2},n_{1}^{\prime},n_{2}^{\prime})\right)} \nonumber \\
&  &+ \frac{1}{2}  
\sum_{\vec{n}\neq (0,n_{2})} \sum_{\vec{n}^{\prime} \neq (0,n_{2}^{\prime})}
\delta{\left(\epsilon-h(n_{1}-\phi,n_{2})\right)} 
\delta{\left(\phi-L_{2}(n_{1},n_{2},n_{1}^{\prime},n_{2}^{\prime})\right)} \nonumber \\
&  & +\frac{1}{2} \sum_{\vec{n} } 
\sum_{\vec{n}^{\prime} \neq(0,n_{2}^{\prime})} 
\delta{\left(\epsilon-h(n_{1}-\phi,n_{2})\right)} 
\delta{\left(\phi-L_{3}(n_{1},n_{2},n_{1}^{\prime},n_{2}^{\prime})\right)} \nonumber \\
&  & +\frac{1}{2} \sum_{\vec{n}\neq (0,n_{2})} \sum_{\vec{n}^{\prime}} 
\delta{\left(\epsilon-h(n_{1}+\phi,n_{2})\right)} \delta{\left(\phi-L_{3}(n_{1},n_{2},n_{1}^{\prime},
n_{2}^{\prime})\right)} \; ,
\end{eqnarray}
where $L_{1},L_{2}$ and $L_{3}$ are the values of the flux determined by
the roots of
\begin{eqnarray}
\label{vinc}
h(n_{1}+\phi,n_{2})-h(n_{1}^{\prime}+\phi,n_{2}^{\prime})&=&0 \;, \nonumber \\
h(n_{1}-\phi,n_{2})-h(n_{1}^{\prime}-\phi,n_{2}^{\prime})&=&0 \;, \\
h(n_{1}-\phi,n_{2})-h(n_{1}^{\prime}+\phi,n_{2}^{\prime})&=&0 \;, \nonumber
\end{eqnarray}
respectively.

Now we employ the Poisson's formula. The first sum in Eq.(~\ref{rhocil})
that takes into account the crossings between energy levels that increase
as a function of $\phi$ (this term includes the crossings between
levels $(n_{1}=0,n_{2})$ and $(n_{1}^{\prime}=0,n_{2})$) can be written as,
\begin{eqnarray}
& & \frac{1}{2}\sum_{m_{1},m{2}}\sum_{m_{1}^{\prime},m_{2}^{\prime}}
\int_{\vec{I}\geq(0,0)} \int_{\vec{I^{\prime}}\geq (0,0)}
\delta{\left(\epsilon-h(I_{1}+\phi,I_{2})\right)} 
\delta{\left(\phi-L(I_{1}+\phi,I_{2},  
I_{1}^{\prime}+\phi,I_{2}^{\prime}\right)} \nonumber \\
& & \times \exp{\left(i 2 \pi (\vec{m} \cdot \vec{I} 
+ \vec{m}^{\prime} \cdot \vec{I}^{\prime} )\right)} \;,
\end{eqnarray}
where we have taken into account that $\alpha_{1}=0$ and $\alpha_{2}=4$. To elimate the $\delta$ functions we employ Eqs.(1.7-1.18)
together with the first equation of Eq.(~\ref{vinc}). We obtain
\begin{eqnarray}
\rho_{c}(\epsilon,\phi,+,+)&=&\frac{\gamma}{2}\sum_{\vec{m}} \sum_{\vec{m}^{\prime}}
\int_{I_{1}\geq 0} \int_{I_{1}^{\prime} \geq 0} dI_{1} dI_{1}^{\prime} \nonumber \\
&  & \times
\frac {\left| I_{1}-I_{1}^{\prime}\right|}
{2 \sqrt{\epsilon-\gamma (I_{1}+\phi)^{2}}\sqrt{\epsilon-\gamma (I_{1}^{\prime}+\phi)^2}} 
\exp{\left( i 2 \pi (m_{1}I_{1}+m_{2}I_{2}+ m_{1}^{\prime}I_{1}^{\prime}+
m_{2}^{\prime}I_{2}^{\prime})\right)} \; .
\end{eqnarray}
Now, we redefine $\bar{I}_{1} \; (\bar{I}_{1}^{\prime})\equiv I_{1}+\phi \; (I_{1}^{\prime}+\phi)$. 
After replacing it follow
\begin{eqnarray}
\label{con++}
\rho_{c}(\epsilon,\phi,+,+)&=&\frac{\gamma}{2}\sum_{\vec{m}} \sum_{\vec{m}^{\prime}}
\int_{I_{1}\geq \phi} \int_{I_{1}^{\prime} \geq \phi} dI_{1} dI_{1}^{\prime} \nonumber \\
&  & \times
\frac {\left| I_{1}-I_{1}^{\prime}\right|}
{2 \sqrt{\epsilon-\gamma I_{1}^{2}}\sqrt{\epsilon-\gamma I_{1}^{2\prime}}} 
\exp{\left(i 2 \pi (m_{1}(I_{1}-\phi)+m_{2}I_{2}+ m_{1}^{\prime}(I_{1}^{\prime}-\phi)
+ m_{2}^{\prime}I_{2}^{\prime})\right)} \; ,
\end{eqnarray}
where we have omitted the bar in the integration variables. 
In the same way, the second sum in Eq.(~\ref{rhocil}) (using the second
equation
of Eq.(~\ref{vinc})) and the third and fourth (using the third equation of
Eq.(~\ref{vinc})) are 
\begin{eqnarray}
\label{con--}
\rho_{c}(\epsilon,\phi,-,-)&=&\frac{\gamma}{2}\sum_{\vec{m}} \sum_{\vec{m}^{\prime}}
\int_{I_{1}\geq -\phi} \int_{I_{1}^{\prime} \geq -\phi} dI_{1} dI_{1}^{\prime} \nonumber \\
&  & \times
\frac {\left| I_{1}-I_{1}^{\prime}\right|}
{2 \sqrt{\epsilon-\gamma I_{1}^{2}}\sqrt{\epsilon-\gamma I_{1}^{2\prime}}} 
\exp{\left(i 2 \pi (m_{1}(I_{1}+\phi)+m_{2}I_{2}+ m_{1}^{\prime}(I_{1}^{\prime}+\phi)
+ m_{2}^{\prime}I_{2}^{\prime})\right)} \; ,
\end{eqnarray}
\begin{eqnarray}
\label{con+-}
\rho_{c}(\epsilon,\phi,+,-)&=&\frac{\gamma}{2}\sum_{\vec{m}} \sum_{\vec{m}^{\prime}}
\int_{I_{1}\geq \phi} \int_{I_{1}^{\prime} \geq -\phi} dI_{1} dI_{1}^{\prime} \nonumber \\
&  & \times
\frac {(I_{1}+I_{1}^{\prime})}
{2 \sqrt{\epsilon-\gamma I_{1}^{2}}\sqrt{\epsilon-\gamma I_{1}^{2\prime}}} 
\exp{\left(i 2 \pi (m_{1}(I_{1}-\phi)+m_{2}I_{2}+ m_{1}^{\prime}(I_{1}^{\prime}+\phi)
+ m_{2}^{\prime}I_{2}^{\prime})\right)} \;,
\end{eqnarray}
\begin{eqnarray}
\label{con-+}
\rho_{c}(\epsilon,\phi,-,+)&=&\frac{\gamma}{2}\sum_{\vec{m}} \sum_{\vec{m}^{\prime}}
\int_{I_{1}\geq -\phi} \int_{I_{1}^{\prime} \geq \phi} dI_{1} dI_{1}^{\prime} \nonumber \\
&  & \times
\frac {( I_{1}+I_{1}^{\prime})}
{2 \sqrt{\epsilon-\gamma I_{1}^{2}}\sqrt{\epsilon-\gamma I_{1}^{2\prime}}} 
\exp{\left(i 2 \pi (m_{1}(I_{1}+\phi)+m_{2}I_{2}+ m_{1}^{\prime}(I_{1}^{\prime}-\phi)
+ m_{2}^{\prime}I_{2}^{\prime})\right)} \; .
\end{eqnarray}
Equation (~\ref{con--}) takes into account the crossings between energy levels that decrease as a function of $\phi$ while Eqs.(~\ref{con+-}) and (~\ref{con-+}) 
correspond to crossings between levels that increase and decrease as a
function of $\phi$.
For the smooth part, we set $m_{1}=m_{2}=m_{1}^{\prime}=m_{2}^{\prime}=0$
in the sum of Eqs.(~\ref{con++}-~\ref{con-+}) and taking into account that the lower limit in the integration over $I_{i}$ can be taken as 0 (because  $\epsilon>>1$ implies $\phi\rightarrow 0^{+}$ and $-\phi \rightarrow 0^{-}$) we obtain Eq.(~\ref{flux}).

To obtain the first contribution to the oscillating part 
$\rho_{c}(\epsilon,\phi)_{osc1}$ Eq.(~\ref{osmain}), we set $m_{1}=m_{2}=0$
and we integrate over $I_{1}$. Then, we approximate the integration over
$I_{1}^{\prime}$ using the stationary phase thechnique obtaining 
Eq.(~\ref{osc1ci}). For the second contribution $\rho_{c}(\epsilon,\phi)_{osc2}$
we evalute both integrals (over $I_{1}$ and $I_{2}$) in the phase stationary
condition. The resulting Eq.(~\ref{osc2ci}) is long but straightforward. However
we will remark some steps. After replacing the phase stationary condition
for the actions and reducing the sums over $m_{i}$ and $m_{i}^{\prime}$
to the positive quadrants we will have in Eqs.(~\ref{con++}) and (~\ref{con--}) terms like (we write the prefactors as $A$ ),
\begin{eqnarray}
\label{desdo1}
&  & 2 A \cos{(S-2\pi m_{1}\phi)-\pi/4} \cos{(S^{\prime}-2\pi m_{1}^{\prime} \phi-\pi/4)} \; ,\nonumber\\
&  & 2 A \cos{(S+2\pi m_{1}\phi)-\pi/4} \cos{(S^{\prime}+2\pi m_{1}^{\prime} \phi-\pi/4)} \; ,
\end{eqnarray}
respectively. These terms can be appropriately combined using trigonometric
identities in the form
\begin{equation}
 A \left[\cos{(S-S^{\prime})}\cos{\left(2\pi \phi(m_{1}-m_{1}^{\prime})\right)}
+ \cos{(S+S^{\prime}+\pi/2)}\cos{\left(2\pi\phi(m_{1}+m_{1}^{\prime}) \right)}\right] \;.
\end{equation}
In the same way the terms of Eqs.(~\ref{con+-}) and (~\ref{con-+}) result   
\begin{eqnarray}
\label{desdo2}
&  & 2 B \cos{(S-2\pi m_{1}\phi-\pi/4)} \cos{(S^{\prime}+2\pi m_{1}^{\prime} \phi-\pi/4)} \;,\nonumber\\
&  & 2 B \cos{(S+2\pi m_{1}\phi-\pi/4)} \cos{(S^{\prime}-2\pi m_{1}^{\prime} \phi-\pi/4)} \; .
\end{eqnarray}
Therefore the sum of these terms is
\begin{equation}
 B \left[\cos{(S-S^{\prime})}\cos{\left(2\pi\phi(m_{1}+m_{1}^{\prime})\right)}
+ \cos{(S+S^{\prime}+\pi/2)}\cos{\left(2\pi\phi(m_{1}-m_{1}^{\prime} )\right)} \right] \;,
\end{equation}
and Eq.(~\ref{osc2ci}) follows.
Let us remark that Eqs.(~\ref{desdo1}) and (~\ref{desdo2}) are a consequence of the
breaking of the time reversal symmetry when the magnetic flux is present. In such a case, twin classical orbits (related to each othe by the time reversal transformation) that have the same action when the systems has the time reversal symmetry split their actions when the flux is applied \cite{bohi2}.

\section*{Acknowledgments}
This work was partially supported by UBACYT (TW35), PICT97 03-00050-01015 and CONICET.\\

\newpage

\newpage

\newpage

\newpage

\newpage

\newpage

\newpage

\end{document}